\documentclass[10pt,conference,twocolumn]{IEEEtran}

\usepackage{amsmath}
\usepackage{amssymb}
\usepackage{latexsym}
\usepackage{amsfonts}

\begin{document}

\title{An Algorithm for Computing  $m$-Tight Error Linear Complexity of Sequences over $GF(p^{m})$ with Period $p^{m}$}

\author{
\authorblockN{ Jianqin Zhou$^{1,2}$,Wei Xiong$^{1}$}
\authorblockA{ 1.Telecommunication School, Hangzhou Dianzi University,
Hangzhou, 310018 China\\
2.Computer Science School, Anhui University of Technology, Maanshan, 243002 China\\
 \ \ zhou9@yahoo.com}
 }

\maketitle

\begin{abstract}The linear complexity (LC) of a sequence has been used as a convenient measure
of the randomness of a sequence.  Based on the theories of linear
complexity, $k$-error linear complexity, the minimum error and the
$k$-error linear complexity profile, the notion of $m$-tight error
linear complexity is presented. An efficient algorithm for computing
$m$-tight error linear complexity is derived from the algorithm for
computing $k$-error linear complexity of sequences over GF($p^{m}$)
with period $p^n$, where $p$ is a prime. The validity of the
algorithm is shown. The algorithm is also realized with C language,
and an example is presented to illustrate the algorithm.

\indent {\bf Keywords:} {\it Period sequence; linear complexity;
k-error linear complexity; tight error linear complexity}

\end{abstract}

\section{Introduction}

Among the measures commonly used to measure the complexity of a
sequence(S) is its linear complexity LC(S), defined as the length of
the shortest linear feedback shift register that generates
sequence(S). According to the Berlekamp-Massey algorithm
\cite{Blackburn,Massey}, if the linear complexity of sequence(S) is
LC(S), and 2LC(S) consecutive elements of the sequence are known,
then we can find the homogeneous linear recurrence  relation of the
sequence by solving linear equations or B-M algorithm, then the
whole sequence is determined. So the linear complexity of key
sequence must be large enough to oppugn known plain text attack.

However, a high linear complexity can not necessarily guarantee the
sequence is safe. For example, the first period of a binary sequence
with period $n$ is $S=\overbrace{0,0,\cdots,0,1}$, its linear
complexity is $n$, but the linear complexity declines to $0$ when
change the last element to $0$. The linear complexity of these
sequences are unstable, and these sequences used as key stream are
unsafe. Therefore, the linear complexity stability of period
sequence is closely related to the unpredictability of the sequence.
Not only the linear complexity of period sequence should be large
enough, but also the linear complexity stability should be high.

Ding, Xiao and Shan \cite{Ding}  first noted this phenomenon and
presented the weight complexity and sphere complexity. Similarly,
Stamp and Martin \cite{Stamp} introduced $k$-error linear
complexity, which is defined to be the smallest linear complexity
that can be obtained when any $k$ or fewer of the symbols of the
sequence are changed within one period, and presented the concept of
$k$-error linear complexity profile. It is known that the sphere
complexity defined by Ding, Xiao, and Shan in \cite{Ding} is earlier
than the $k$-$LC$ and they are essentially the same (but not
completely the same).

 The $k$-error linear complexity of any sequence can be also
calculated by using B-M algorithm repeatedly. But in order to
compute the $k$-error linear complexity of binary sequences with
period $N$, this algorithm must be used $\sum\limits_{j=0}^{k}\left(
\begin{array}{c}N\\j\end{array}\right)$ times. For binary sequences with period $N$,
although we had some algorithms for determining linear complexity of
particular period sequences, if we do not have an effective
algorithm to compute the $k$-error linear complexity, fast algorithm
also should be used $\sum\limits_{j=0}^{k}\left(
\begin{array}{c}N\\j\end{array}\right)$ times. Even $N$ and $k$ is not large enough,
the computation is still considerable.

Based on Games-Chan  algorithm \cite{Games}, Stamp and Martin
\cite{Stamp} presented a fast algorithm for determining $k$-error
linear complexity of binary sequence with period $2^{n}$. By using
the modified cost different from that used in the Stamp-Martin
algorithm for sequences over GF(2) with period $2^{n}$, Kaida,
Uehara and Imamura \cite{Kaida}  presented a fast algorithm for
determining the $k$-error linear complexity of  sequences with
period $p^{n}$ over GF($p^{m}$), p a prime.

The reason why people study the stability of linear complexity is
that a small number of changes may lead to a sharp decline of linear
complexity. How many elements have to be changed to reduce the
linear complexity? Kurosawa et al. \cite{Kurosawa} introduced the
concept of minerror($S$) to deal with the problem, and defined it as
the least  number $k$ for which the $k$-error linear complexity is
strictly less than the linear complexity, which is corresponding to
the $k$-value of the first jump point of  $k$-error linear
complexity profile.

What is the linear complexity  after decline? That is, what is the
value of $k$-error linear complexity when $k$ = minerror($S$)?
Aiming at these problems, the relation between the linear complexity
 and $k$-error linear
complexity of binary sequences with period $2^n$ is studied in
\cite{Kurosawa}, the minerror($S$) denoted by the Hamming weight of
linear complexity  is given, and the upper bound of $k$-error linear
complexity for $k$ = minerror($S$) is also given. Meidl \cite{Meidl}
studied the stability of linear complexity of binary sequence with
period $p^{n}$, and proved the upper and lower bound of minerror(S).

 The error linear complexity spectrum of a periodic sequence is
 introduced by Lauder
and Paterson \cite{Lauder}
 to indicate how
linear complexity decreases as the number $k$ of bits allowed to be
modified per period increases, the same as $k$-error linear
complexity profile  defined in \cite{Stamp}. Moreover, Lauder and
Paterson \cite{Lauder} generalized the algorithm in \cite{Stamp} to
compute the entire error linear complexity spectrum of such
sequences.

In this paper, based on linear complexity, $k$-error linear
complexity, $k$-error linear complexity profile and minerror(S), the
$m$-tight error linear complexity is presented to study the
stability of the linear complexity of periodic sequences. The
$m$-tight error linear complexity is defined as a two tuple
$(k_m,C_m)$, which is the $m$th jump point of the $k$-error linear
complexity profile of a sequence.

A fast algorithm is proposed for determining the $m$-tight error
linear complexity of sequences over GF($p^{m}$) with period $p^{n}$,
where $p$ is a prime. The algorithm is derived from the algorithm
for the k-error linear complexity of sequences over GF($p^{m}$) with
period $p^{n}$, where $p$ is a prime \cite{Kaida}. The proposed
algorithm is realized with C language, and an example is presented
to illustrate the algorithm.

The paper is organized as follows. Section II introduces $k$-error
linear complexity  algorithm presented by Kaida, Uehara and Imamura
\cite{Kaida},
 whereas Section III
 focuses on the algorithm for determining the $m$-tight error
linear complexity of sequences. Concluding remarks are given in
Section IV.

\section{$k$-error linear complexity  algorithm }

  In this paper we will consider sequences over $GF(q)$ with period $p^n, n\ge1$, where
$q=p^m$ and $p$ is a prime. In the following algorithms,
$\overrightarrow{X}$ denotes a vector.

Algorithm 2.1 is got by generalizing Games-Chan algorithm
\cite{Ding,Kaida}. Let $\{a_i\}=\{a_{0} , a_{1} ,a_{2} ,\ldots\}$ be
a sequence with period $N=p^{n}$ over $GF(q)$, where $q=p^{m}$, $p$
is a prime number. Let
$\overrightarrow{a}^{(N)}=(a_{0}^{(N)},a_{1}^{(N)},\cdots,a_{N-1}^{(N)})$
be the first period of the sequence. It is divided  into $p$ parts
and denoted as
$\overrightarrow{a}^{(N)}=(a(0)^{(N)},\cdots,a(p-1)^{(N)})$, where
$\overrightarrow{a}(j)^{(N)}=(a_{jM}^{(N)},\cdots,a_{(j+1)M-1}^{(N)})$.

\

{\bf Algorithm 2.1 Generalized Games-Chan algorithm}

//Initial values:

$N=pM,LC=0,q=p^m$,

$\overrightarrow{a}^{(N)}=(a_{0}^{(N)},a_{1}^{(N)},\cdots,a_{N-1}^{(N)})$

While $M>1$ do

\indent \indent
$\overrightarrow{a}^{(pM)}=(a(0)^{(pM)},\cdots,a(p-1)^{(pM)})$

\indent \indent for $u=0,\cdots,p-1$ do\\
\indent \indent \indent  \begin{eqnarray*}
&&\overrightarrow{b}(u)^{(M)}\\
&=&F_u(\overrightarrow{a}(0)^{(pM)},\cdots,\overrightarrow{a}(p-1)^{(pM)})\\
&=&\sum\limits_{j=0}^{p-u-1}c_{u,j}\overrightarrow{a}(j)^{(pM)}\\
&=&\sum\limits_{j=0}^{p-u-1}\left(
\begin{array}{c}{p-j-1}\\u\end{array}\right)
\overrightarrow{a}(j)^{(pM)}\end{eqnarray*}

\indent \indent end for

\indent \indent if
$\overrightarrow{b}(0)^{M}=\cdots=\overrightarrow{b}(p-1)^{M}=\overrightarrow{0}$
then

\indent \indent \indent $w=1$

\indent \indent end if

\indent \indent for $w_1=2,\cdots,p-1$  do

\indent \indent \indent if
$\overrightarrow{b}(0)^{(M)}=\cdots=\overrightarrow{b}(p-w_1-1)^{(M)}=\overrightarrow{0}$

\indent \indent \indent and
$\overrightarrow{b}(p-w_1)^{(M)}\neq\overrightarrow{0}$ then

\indent \indent \indent \indent $w=w_1$

\indent \indent \indent end if

\indent \indent end for

\indent \indent if
$\overrightarrow{b}(0)^{(M)}\neq\overrightarrow{0}$ then

\indent \indent \indent $w=p$

\indent \indent end if

\indent \indent
$\overrightarrow{a}^{(M)}=F_{p-w}(\overrightarrow{a}(0)^{(pM)},\cdots,\overrightarrow{a}(p-1)^{(pM)})$

\indent \indent $LC=LC+(w-1)M$

\indent \indent $M=M/p$

end while

$\overrightarrow{a}^{(1)}=(a_0^{(1)})$

if $a_0^{(1)}\neq0$ then

\indent \indent $LC=LC+1$

end if\\

Using  Games-Chan algorithm,  Stamp-Martin algorithm  \cite{Stamp}
 computes the $k$-$LC$ of sequences over $GF(2)$ with period $2^n$.
Algorithm 2.2 is got by using generalized Games-Chan algorithm
\cite{Kaida}.

The cost of $\overrightarrow{a}^{(M)}$ is $AC(M)$, which is a
$q\times{M}$ matrix. Further define the matrix as
$AC(M)=[A(h,i)_M]$, where $A(h,i)_M$ is the minimum number of
changes required in the original sequence $\overrightarrow{a}^{(N)}$
to change the current element $\alpha_i^{(M)}$ to
$\alpha_i^{(M)}+\partial_h$.
 The cost of
$\overrightarrow{b}(u)^{(M)}$ is $BC(M)$ , which is a
$(p-1)\times{M}$ matrix. Further define the matrix as
$BC(M)=[B(u,i)_M]$, where $B(u,i)_M$ is  the minimum number of
changes required in the original sequence $\overrightarrow{a}^{(N)}$
to force
 $b_{0,i}^{(M)}=\cdots=b_{u,i}^{(M)}=0$.

 \

{\bf Algorithm 2.2 Kaida-Uehara-Imamura algorithm}

//Initial values:

$N=pM=p^n,k\mbox{-}LC=0,$

$\overrightarrow{a}^{(N)}=(\overrightarrow{a}_0^{(N)},\overrightarrow{a}_1^{(N)},\cdots,
\overrightarrow{a}_{N-1}^{(N)}),q=p^m$

 for
$h=0,1,\cdots,q-1,i=0,1,\cdots,N-1$ do

\indent \indent $AC(N)=[A(h,i)_N]=\left\{\begin{array}{l}
0, \mbox{if } h=0,\ \   \\
1, \mbox{if } h\neq0. \ \
\end{array}\right.$

while $M>1$ do\\
\indent \indent for $u=0,1,\cdots,p-2,i=0,1,\cdots,M-1$ do \\
\indent \indent \indent{\scriptsize
$B(u,i)_M=\min\{\sum\limits_{j=0}^{p-1}A(e_j,i+jM)pM|\overrightarrow{e}\in{D(u,i)_M\}}$}\\
\indent \indent \indent where
$\overrightarrow{e}=(e_0,\cdots,e_{p-1})\in[GF(q)]^p$ and\\
\indent \indent \indent {\scriptsize
$D(u,i)_M=\{\overrightarrow{e}|F_j(e_0,\ldots,e_{p-1})+b_{j,i}^{(M)}=0(0\leq{j}\leq{u})\}$}\\
\indent \indent \indent $TB(u)_M=\sum\limits_{i=0}^{M-1}B(u,i)_M$

\indent \indent end for

\indent \indent if $TB(p-2)_M\leq{k}$,then\\
\indent \indent \indent $w=1$\\
\indent \indent end if\\
\indent \indent for $w_1=2,\cdots,p-1$ do\\
\indent \indent \indent if {\scriptsize
$TB(p-w_1-1)_M\leq{k}<TB(p-w_1)_M$} then\\
\indent \indent \indent \indent $w=w_1$\\
\indent \indent \indent end if

\indent \indent end for

\indent \indent if $k<TB(0)_M$, then\\
\indent \indent \indent $w=p$\\
\indent \indent end if\\
\indent \indent $\overrightarrow{a}=F_{p-w}(\overrightarrow{a}{(0)}^{(pM)},\cdots,\overrightarrow{a}{(p-1)}^{(pM)})$\\
\indent \indent $k\mbox{-}LC=k\mbox{-}LC+(w-1)M$\\
\indent \indent for $h=0,1,\cdots,q-1,i=0,1,\cdots,M-1$ do\\
\indent \indent \indent {\scriptsize
$A(h,i)_M=\min\{\sum\limits_{j=0}^{p-1}A(e_j,i+jM)pM|\overrightarrow{e}\in{\widehat{D}(u,i)_M^w\}}$}\\

\indent \indent \indent where

\indent \indent \indent {\scriptsize
$\widehat{D}(h,i)_M^1\\
\indent \indent \indent=\left\{\overrightarrow{e}|\begin{array}{l}
F_j(e_0,\cdots,e_{p-1})+b_{j,i}^M=0(0\leq{j}\leq{p-2}),\ \  \\
e_0-\partial_h=0,\ \
\end{array}\right\},$}\\
\indent \indent \indent for $w=1$;\\
\indent \indent \indent {\scriptsize $\widehat{D}(h,i)_M^w\\
 \indent
\indent \indent=\left\{\overrightarrow{e}|\begin{array}{l}
F_j(e_0,\cdots,e_{p-1})+b_{j,i}^M=0(0\leq{j}\leq{p-w-1}),\ \  \\
F_{p-w}(e_0,\cdots,e_{p-1})-\partial_h=0,\ \
\end{array}\right\},$}\\
\indent \indent \indent for $2\leq{w}\leq{p-1}$;\\
\indent \indent \indent
$\widehat{D}(h,i)_M^p=\{\overrightarrow{e}|F_0(e_0,\cdots,e_{p-1})-\partial{h}=0\},$\\
\indent \indent \indent for $w=p$;

\indent \indent end for

\indent \indent $M=M/p$

end while

$\overrightarrow{a}^{(1)}=(a_0^{(1)}),AC(1)=[A(h,0)_1]$

if $A(-a_0^{(1)},0)_1>k$ then

\indent \indent $k\mbox{-}LC=k\mbox{-}LC+1$

 end if

\section{$m$-tight error linear complexity algorithm }

The $m$-tight error linear complexity of sequence S is defined to be
a two tuple $(k_m,C_m)$, which is the $m$th jump point of the
$k$-error linear complexity profile of sequence S. Obviously,
$0$-tight error linear complexity is $(0,C_0)$, $C_0$ is the linear
complexity. In the case of $1$-tight error linear complexity
$(k_1,C_1)$, $k_1$ is the least number to force linear complexity
decline, which is the minerror(S)  defined by Kurosawa et al., and
$C_1$ is $k_1$-error linear complexity.

Based on Algorithm 2.2, it is easy to compute $m$-tight error linear
complexity of sequences over GF($p^m$) with period $p^n$. Firstly,
algorithm 2.2 is changed as follows:

Before while loop add

$T_{min}=N$;

Before $k$-$LC=k$-$LC+(w-1)M$, add

if $TB[p-2]_M>k$ and $TB[p-w]_M<T_{min}$ then $T_{min}=TB[p-w]_M$;

Before $k$-$LC=k$-$LC+1$, add

if $A(-a_0^{(1)},0)_1<T_{min}$ then $T_{min}=A(-a_0^{(1)},0)_1$.

The modified algorithm is denoted as \textbf{Algorithm 3.1}. First
call Algorithm 3.1 with $k=0$, we get $0$-error linear complexity
$c_0$ of original sequence, so $0$-tight error linear complexity is
$(0,c_0)$. Meanwhile we get $T_{min}$, denoted as  $k_1$. Call
Algorithm 3.1 with $k=k_1$, we get $k_1$-error linear complexity
$(k_1,c_1)$ of original sequence, meanwhile we obtain $T_{min}$,
denoted as $k_2$. Call Algorithm 3.1 with $k=k_2$, we get
$k_2$-error linear complexity of original sequence, that is
$2$-tight error linear complexity is $(k_2,c_2)$. Meanwhile we
obtain $T_{min}$, denoted as $k_3$. Call Algorithm 3.1 recursively,
we can obtain $m$-tight error linear complexity $(k_m,c_m)$ of
original sequence.

Algorithm 3.1 starts the recursive process from $0$-error linear
complexity. While  compute \emph{k}-error linear complexity, we also
 compute
 the minimum number $T_{min}$  of
changes required in the original sequence to force \emph{k}-error
linear complexity to decline.

In \cite{Kaida}, $TB(u)_M$ is defined as the minimum number of
changes in $a^{(N)}$ necessary and sufficient for making

$b(0)^{(M)}=\cdots=b(u)^{(M)}=0$, $0\le u\le p-2$.

In the process of computing \emph{k}-error linear complexity, we
must try to force  $TB(p-w)_M\leq{k}, w\ge2$ or
$A(-a_0^{(1)},0)_1\leq{k}$. Thus, the minimum number $T_{min}$  of
changes required in the original sequence to force \emph{k}-error
linear complexity to decline is the smallest of  those $TB(p-w)_M,
w\ge2$ or $A(-a_0^{(1)},0)_1$.

Therefore the validity of our algorithm is shown.

We now compute the tight error linear complexity of sequence $S$ by
Algorithm 3.1. Let $S$ be a sequence with period $N=p^n$, the first
period of $S$ is
$S^{27}=0,2,0,2,1,1,0,1,0,1,2,0,1,1,1,0,1,0,2,2,0,2,1,1,0,1,0.$

Apply Algorithm 3.1, we get the following results:

The first step,$k=0$:

$M=9: TB[0]=1,TB[1]=3,w=3,k$-$LC=18$;

$M=3: TB[0]=1,TB[1]=1,w=3,k$-$LC=24$;

$M=1: TB[0]=1,TB[1]=1,w=3,k$-$LC=26$;

$k$-$LC=27,T_{min}=1$.

The second step,$k=1$:

$M=9: TB[0]=1,TB[1]=3,w=2,k$-$LC=9$;

$M=3: TB[0]=1,TB[1]=4,w=2,k$-$LC=12$;

$M=1: TB[0]=4,TB[1]=4,w=3,k$-$LC=14$;

$k$-$LC=15, T_{min}=3$.

The third step,$k=3$:

$M=9: TB[0]=1,TB[1]=3,w=1,k$-$LC=0$;

$M=3: TB[0]=9,TB[1]=11,w=3,k$-$LC=6$;

$M=1: TB[0]=3,TB[1]=3,w=1,k$-$LC=6$;

$k$-$LC=7, T_{min}=9$.

The fourth step,$k=9$:

$M=9: TB[0]=1,TB[1]=3,w=1,k$-$LC=0$;

$M=3: TB[0]=9,TB[1]=11,w=2,k$-$LC=3$;

$M=1: TB[0]=10,TB[1]=10,w=3,k$-$LC=5$;

$k$-$LC=6, T_{min}=10$.

The fifth step,$k=10$:

$M=9: TB[0]=1,TB[1]=3,w=1,k$-$LC=0$;

$M=3: TB[0]=9,TB[1]=11,w=2,k$-$LC=3$;

$M=1: TB[0]=10,TB[1]=10,w=1,k$-$LC=3$;

$k$-$LC=4, T_{min}=11$.

The sixth step,$k=11$:

$M=9: TB[0]=1,TB[1]=3,w=1,k$-$LC=0$;

$M=3: TB[0]=9,TB[1]=11,w=1,k$-$LC=0$;

$M=1: TB[0]=12,TB[1]=16,w=3,k$-$LC=2$;

$k$-$LC=3, T_{min}=12$.

The seventh step,$k=12$:

$M=9: TB[0]=1,TB[1]=3,w=1,k$-$LC=0$;

$M=3: TB[0]=9,TB[1]=11,w=1,k$-$LC=0$;

$M=1: TB[0]=12,TB[1]=16,w=2,k$-$LC=1$;

$k$-$LC=2, T_{min}=16$.

The eighth step,$k=16$:

$M=9: TB[0]=1,TB[1]=3,w=1,k$-$LC=0$;

$M=3: TB[0]=9,TB[1]=11,w=1,k$-$LC=0$;

$M=1: TB[0]=12,TB[1]=16,w=1,k$-$LC=0$;

$k$-$LC=1$, $T_{min}=17$ .

The ninth step,$k=17$:

$M=9: TB[0]=1,TB[1]=3,w=1,k$-$LC=0$;

$M=3: TB[0]=9,TB[1]=11,w=1,k$-$LC=0$;

$M=1: TB[0]=12,TB[1]=16,w=1,k$-$LC=0$;

$k$-$LC=0$.

By calling algorithm 3.1, the tight error linear  complexity is
obtained successively: (0,27), (1,15), (3,7), (9,6), (10,4),
(11,3),(12,2), (16,1),(17,0).

\section{Conclusion}

Lauder and Paterson \cite{Lauder} presented an algorithm to compute
the error linear complexity spectrum of a binary sequence of period
$2^n$. However, our algorithm is more suitable to compute
minerror(S) or $m$-tight error linear complexity for small $m$.

Based on relevant theoretical basis of $k$-error linear complexity,
we proposed $m$-tight error linear complexity to study the stability
of stream cipher. Based on Kaida-Uehara-Imamura algorithm,  we
presented  a fast algorithm for determining the $m$-tight error
linear complexity of sequences over GF($p^m$) with period $p^n$,
where $p$ is a prime number. The concept of $m$-tight error linear
complexity integrates all linear complexity, $k$-error linear
complexity,   $k$-error linear complexity profile and the concept of
minerror(S). So the fast algorithm for determining $m$-tight error
linear complexity has important theoretical significance and
application value.

\section*{Acknowledgment}
The research was supported by  Zhejiang Natural Science Foundation
(No. Y1100318, R1090138), Opening Project of Shanghai Key Laboratory
of Integrate Administration Technologies for Information Security.

{\bf Jianqin Zhou} received his B.Sc. degree in mathematics from
East China Normal University, China, in 1983, and M.Sc. degree in
probability and statistics from Fudan University, China, in 1989.
From 1989 to 1999 he was with the Department of Mathematics and
Computer Science, Qufu Normal University, China. From 2000 to 2002,
he worked for a number of IT companies in Japan. From 2003 to 2007
he was with the Department of Computer Science, Anhui University of
Technology, China.
 From Sep 2006  to
Feb 2007, he was a visiting scholar with the Department of
Information and Computer Science, Keio University, Japan. Since 2008
he has been with the Telecommunication School, Hangzhou Dianzi
University, China

He  published more than  70 papers, and  proved a conjecture posed
by famous mathematician Paul Erd\H{o}s et al. His research interests
include cryptography, coding theory  and combinatorics.

{\bf Wei Xiong} received her B.Eng. degree in Electronic Information
Engineering from Xianyang Normal College, China, in 2008,and M.Eng.
degree  in Signal and Information Processing from Hangzhou Dianzi
University, China, in 2011. Her research interests include
cryptography and digital watermark.

\end{document}